\documentclass[aip,pof]{revtex4-1}
\usepackage[pdftex,bookmarks,colorlinks]{hyperref}
\pdfoutput=1

\usepackage[pdftex]{graphicx}
\usepackage{natbib}
\usepackage[pdftex]{graphicx}
\usepackage{bm}
\usepackage{amssymb}
\usepackage{amsbsy}
\usepackage{color}
\usepackage{mathrsfs,mathcomp}
\usepackage{amsmath}
\usepackage{amsfonts}
\usepackage{amssymb}
\usepackage{amsbsy}
\usepackage{color}
\usepackage{mathrsfs}
\usepackage{graphicx}
\usepackage{subfigure}
\usepackage{multirow}

\newcommand{\be}{\begin{equation}}
\newcommand{\ee}{\end{equation}}
\newcommand{\bea}{\begin{eqnarray}}
\newcommand{\eea}{\end{eqnarray}}
\newcommand{\pat}{\partial}

\newcommand{\lam}{\lambda}

\begin{document}

\title{Oscillations of a gas pocket on a liquid-covered solid surface}
\author{Hanneke Gelderblom}
 \email{h.gelderblom@tnw.utwente.nl.}
\author{Aaldert G. Zijlstra}%
\author{Leen van Wijngaarden}
\affiliation{ 
Physics of Fluids Group, Faculty of Science and Technology, J. M. Burgers Centre for Fluid Dynamics, University of Twente, 7500 AE Enschede, The Netherlands}
\author{Andrea Prosperetti}
\affiliation{ 
Physics of Fluids Group, Faculty of Science and Technology, J. M. Burgers Centre for Fluid Dynamics, University of Twente, 7500 AE Enschede, The Netherlands}
 \affiliation{Department of Mechanical Engineering, Johns Hopkins University, Baltimore, MD 21218, USA.}

\date{\today}

\begin{abstract}
The dynamic response of a gas bubble entrapped in a cavity on the surface of a submerged solid subject to an acoustic field is investigated in the linear approximation. We derive semi-analytical expressions for the resonance frequency, damping and interface shape of the bubble. For the liquid phase, we consider two limit cases: potential flow and unsteady Stokes flow. The oscillation frequency and interface shape are found to depend on two dimensionless parameters: the ratio of the gas stiffness to the surface tension stiffness, and the Ohnesorge number, representing the relative importance of viscous forces. We perform a parametric study and show, among others, that an increase in the gas pressure or a decrease in the surface tension leads to an increase in the resonance frequency until an asymptotic value is reached. 
\end{abstract}

\maketitle
\section{Introduction}\label{introduction}
The volume pulsations of a gas pocket entrapped on a liquid-covered 
solid surface constitute a fundamental problem at the root of several 
applications in biology, microfluidics, cavitation and others. For example, the 
oscillatory flow induced by the pulsations causes a liquid motion which can be 
used to study the behavior of bacteria and cells under the action of 
shear~\cite[see e.g.][]{Miller:1998, Kuznetsova:2005, Zinin:2009}. In these 
conditions sonoporation of cell walls may occur, which would facilitate the 
uptake of drugs~\cite[see e.g.][]{Postema:2007} and gene 
transfection~\cite[see e.g.][]{Browning:2012}. The flow induced by the 
oscillating gas pocket also induces mixing and 
streaming~\cite[see e.g.][]{Liu:2002}. Under large-amplitude acoustic 
excitation, small gas bubbles issue from the gas entrapped in the cavities 
which greatly enhance sonochemical reactions in a more controlled 
way than is possible in a conventional sonoreactor~\cite{Rivas:2012}. 
Microfabricated cavities on a silicon surface have been used to study
controlled cavitation and bubble growth and 
collapse~\cite{Bremond:2006a, Borkent:2009}.

Despite this wide range of applications, little is known about the dynamic response of a 
gas pocket on a submerged solid in an acoustic field.
Miller~\cite{Miller:1982} and Neppiras \emph{et al.}~\cite{Neppiras:1983} recorded the acoustic response of multiple bubbles 
entrapped in a membrane. However, their size was not controlled, and no 
information about the response of the individual bubbles could be obtained. 
Rathgen \emph{et al.}~\cite{Rathgen:2007} studied the dynamics of periodic arrays of 
gas-filled micropores of controlled size on a solid surface.  
Using optical diffraction techniques, they were able to resolve in time, with a high 
accuracy, the nanometer-scale oscillations of the gas-liquid menisci 
driven by a sound field. However, they were unable to resolve the shape of 
the menisci in the course of the oscillations. 

Theoretical studies mainly focused on spherical bubbles in the bulk liquid~\cite{Plesset:1977}  whereas, for crevice bubbles, only approximate results exist. 
Miller \& Nyborg~\cite{Miller:1983} derived approximate expressions for the lowest resonance
frequency and damping of a gas-filled pore on a solid surface under the assumption 
that the interface shape is parabolic. Their result is that the lowest resonance frequency $f_0$ of a cylindrical pore with radius $a$ and depth $h$ is approximately given by
\begin{equation}
f_0=\frac{1}{2\pi a}\sqrt{\frac{15\pi\kappa \lambda p_0a+120\pi \sigma}{32\rho a}}, \label{miller}
\end{equation}
with $\kappa$ the polytropic index, $\lambda=a/h$ the aspect ratio of the pore, $p_0$ the gas pressure when the interface is flat, $\sigma$ the surface tension coefficient, and $\rho$ the liquid density. As an example, upon taking $\lambda=1$, this relation predicts natural frequencies of 176, 17.6, and 1.76 kHz for gas pockets of air in water with equivalent spherical radii of 10, 100, and 1000 $\mu$m. 
To obtain (\ref{miller}), an energy argument was used. 
Rathgen \emph{et al.}~\cite{Rathgen:2007} improved somewhat on this estimate 
by formulating the correct hydrodynamic problem for the 
liquid phase. However, they only solved the problem in an approximate way, retaining the parabolic approximation for the free-surface shape
and also considering only the lowest resonance frequency.

The purpose of the present work is to study the dynamics of the 
liquid-gas interface bounding the gas contained in a cavity at the 
surface of a solid in the linear approximation. We calculate the frequency, damping and surface shape 
of the linear normal modes of oscillation of the system in the inviscid and 
viscous cases. In many situations the resonance frequency of the system is mainly determined by the inertia of the liquid, and hence can be calculated with sufficient accuracy from
a potential flow model.  We estimate the damping in two ways: from the potential flow solution by using a dissipation function 
method and, more accurately, by solving the time-dependent Stokes
equations. The dynamics of the liquid-gas interface is found to depend on two dimensionless parameters: the ratio of the gas stiffness to the surface tension stiffness, and the Ohnesorge number, representing the viscous damping during one period of oscillation. 

\section{Problem formulation}
\begin{figure}
\begin{center}
\includegraphics[width=4 in]{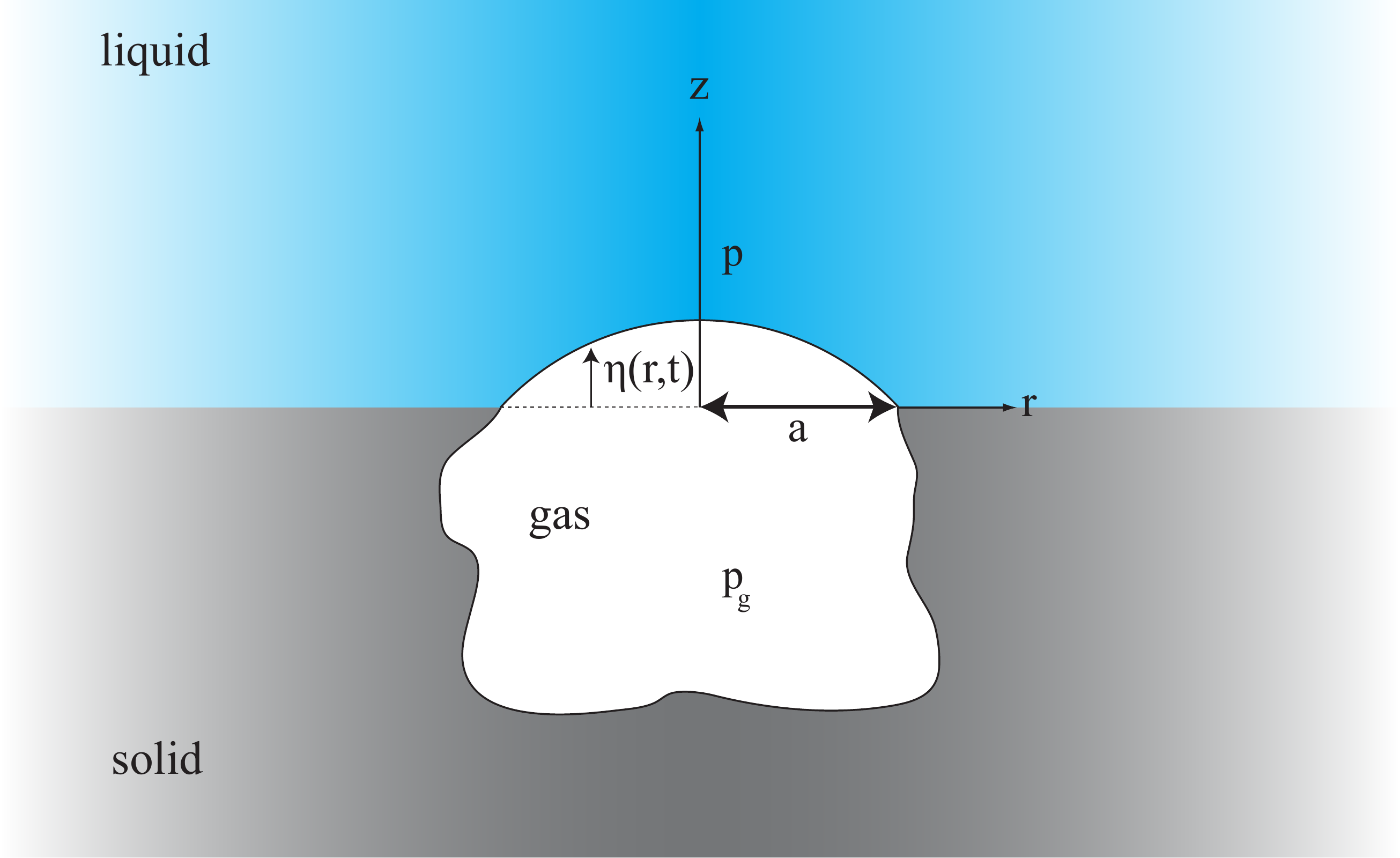}
\caption{A cavity with an entrapped gas bubble. The radius of the circular cavity mouth $a$ is indicated, as well as the cylindrical coordinate system $(r,z)$. The pressures in the gas and in the liquid are $p_g$ and $p$, respectively. The elevation of the free surface above the $(z=0)$-plane is denoted by $\eta(r,t)$. \label{probsketch}}
\end{center}
\end{figure}

Our aim is to describe the resonance frequency and interface shape of a gas bubble entrapped in a crevice. We model the crevice as a cavity with a circular mouth at the surface of an infinite solid submerged in an incompressible liquid (Fig. \ref{probsketch}).  The cavity has an aspect ratio 
\begin{equation}
\lam=\pi a^3/V_0, 
\end{equation}
with $a$ the mouth radius and $V_0$ the cavity volume when the interface is flat. For a cylindrical cavity, $\lam=a/h$ with $h$ the depth of the cavity, but the results that follow hold for cavities of arbitrary shape. We introduce a cylindrical coordinate system $(r, z)$, with the origin located on the axis  of the cavity mouth at the level of the infinite solid plane. The liquid-gas interface is assumed to remain pinned at the circular edge of the cavity. The elevation of the free surface over the plane $z=0$ is described by $\eta(r,t)$, and is assumed to be small compared to the radius of the cavity mouth, $\eta<<a$. 

When the interface is perturbed, the compression of the gas and the surface tension of the liquid-air interface provide restoring forces and the interface will start to oscillate around its equilibrium position, which is assumed to be flat. The oscillating interface causes a velocity field $\boldsymbol{u}$ and pressure field $p$ in the liquid phase (the gas flow is neglected here). To calculate the oscillation frequency, damping and shape of the interface we need to couple the normal stress in the liquid, derived from the velocity field, to the pressure in the gas, which results from the effect of surface tension and the gas compression and expansion. 

We use a linear theory, in which the time-dependence is assumed to be proportional to $e^{\zeta t}$, so that $\eta(r,t)=\eta(r) e^{\zeta t}$, with complex eigenfrequency $\zeta=i \omega-\beta$, where $\omega=2\pi f$ is the angular frequency and $\beta$ the damping coefficient due to viscous dissipation in the liquid. This time-dependence is left implicit in the expressions that follow. 
The motion of the liquid-air interface, described by $S(r,t)=z-\eta(r,t)=0$,  is coupled to the velocity field in the liquid via the kinematic condition
\begin{equation}
\frac{\partial S}{\partial t}+ \boldsymbol{u}\cdot \nabla S=0.
\end{equation}
Since we consider a small interface deformation $\eta$, we neglect all terms which are of second order and higher, which leads to a kinematic boundary condition in the form
\begin{equation}
\left.u_z\right|_{z=0}=\zeta \eta,\quad 0\leq r/a<1.\label{bcp2}
\end{equation}
The solid is impermeable for the liquid, and therefore we impose along the remainder of the $(z=0)$-plane
\begin{equation}
\left.u_z\right|_{z=0}=0, \quad 1<r/a<\infty. \label{bcp1}
\end{equation}
Furthermore, as there is no slip on the solid surface, 
\begin{equation}
\left.u_r\right|_{z=0}=0, \quad 1<r/a<\infty. \label{bcp3a}
\end{equation}
On a clean liquid-gas interface, a no-shear-stress boundary condition applies, since the gas viscosity is much smaller than the liquid viscosity: 
\begin{equation}
\tau_{rz}=\mu \left(\frac{\partial u_r}{\partial z}+\frac{\partial u_z}{\partial r}\right)=0,\quad 0\leq r/a<1,\label{bcp3b}
\end{equation}
with $\mu$ the dynamic viscosity.
The presence of impurities modifies the interfacial behavior and may be modeled by a no-slip condition~\cite{Cuenot:1997} 
\begin{equation}
\left.u_r\right|_{z=0}=0,\quad 0\leq r/a<1.\label{bc3}
\end{equation}
Since the mixed boundary value problem (\ref{bcp3a}), (\ref{bcp3b}) leads to a mathematical problem which does not appear to be solvable, we impose the no-slip conditions (\ref{bcp3a}), (\ref{bc3}) on the entire surface in the following analysis.

The coupling of the pressures in the liquid and the gas occurs via the dynamic boundary condition at $z=0$
\begin{equation}
p_g=p+\sigma \mathcal{C}-2\left.\mu \frac{\partial u_z}{\partial z}\right|_{z=0},\label{bcp4a}
\end{equation}
with $p_g$ the pressure in the gas bubble and $\mathcal{C}=-\left(\partial_r^2\eta+\partial_r\eta/r\right)$ the curvature of the free surface; the last term on the right-hand side is the viscous normal stress on the interface. A general relation between the gas volume $V$ and the gas pressure is given by the polytropic expression
\begin{equation}
\frac{p_g}{p_0}=\left(\frac{V_0}{V}\right)^\kappa,\label{poly}
\end{equation}
with $\kappa$ the polytropic index; $\kappa=1$ is applicable to isothermal conditions. 
When the interface is flat ($\eta=0$), $V=V_0$ and $p_g=p_0$, the ambient pressure in the liquid. By expanding (\ref{poly}) for small interface deformations, we find
\begin{equation}
p_g\approx p_0 \left[1-\kappa\left(\frac{V}{V_0}-1\right)\right].\label{pexpan}
\end{equation}
The gas volume can be found from the interface shape by integration:
\begin{equation}
V=V_0+2\pi \int_0^a \eta r \mathrm{d}r=V_0\left[1+\lam\frac{H}{a^3} \right],\label{vbub}
\end{equation}
with 
\begin{equation}
H=2\int_0^a r\eta(r)\mathrm{d}r,\label{eqH}
\end{equation}
proportional to the volume change of the gas due to the interface deformation.
Combining (\ref{pexpan}) and (\ref{vbub}), we obtain 
\begin{equation}
p_g=p_0\left(1- \kappa\lam\frac{H}{a^3} \right),
\end{equation}
and the dynamic boundary condition (\ref{bcp4a}) becomes
\begin{equation}
p_0\left(1-\kappa\lambda\frac{H}{a^3} \right)=p-\sigma \left(\partial_r^2\eta+\frac{1}{r}\partial_r\eta\right)-2\left.\mu \frac{\partial u_z}{\partial z}\right|_{z=0}.\label{bc4}
\end{equation}
In the following sections, the pressure in the liquid will be calculated in two limit cases: potential flow on one hand, as described in Section \ref{potflow}, and unsteady Stokes flow on the other, in Section \ref{stokesflow}. In addition, an estimate of the viscous damping is obtained using a modified potential flow model, where we calculate the dissipation in the bulk from the potential-flow solution (Section \ref{cpotflow}).
Before we proceed, we introduce the following dimensionless quantities 
\begin{eqnarray}
\hat{r}=\frac{r}{a},~\hat{z}=\frac{z}{a},~\hat{\eta}=\frac{\eta}{a},~\hat{u}=\sqrt{\frac{a\rho}{\sigma}}u,~\hat{\zeta}=\sqrt{\frac{a^3\rho}{\sigma}}\zeta, ~\hat{p}=\frac{a}{\sigma}p, ~\hat{H}=\frac{H}{a^3},\label{dimpar}
\end{eqnarray}
which we will use from now on, thereby dropping the carets. 

\section{Potential flow}\label{potflow}
To calculate the liquid pressure used in the dynamic boundary condition (\ref{bc4}) we first neglect the influence of viscosity completely. Then, the flow is irrotational. As the resonance frequency of the system is mainly determined by the inertia in the liquid, it can be obtained from an inviscid flow model with a fair accuracy, as will be seen later.
In the framework of potential flow theory, the no-slip condition (\ref{bcp3a}) at the solid substrate in combination with either (\ref{bcp3b}) or (\ref{bc3}) cannot be enforced.

\subsection{Governing equations}
For irrotational flow, the velocity field can be written in terms of the potential $\phi$ as $
\boldsymbol{u}=\nabla \phi$,
which satisfies the Laplace equation 
$\nabla^2 \phi=0$.
At the liquid-gas interface $\phi$ has to satisfy the kinematic boundary condition (\ref{bcp2})
\begin{equation}
\left.u_z\right|_{z=0}=\left.\partial_z \phi\right|_{z=0}=\zeta \eta,\quad 0\leq r<1,\label{bc2}
\end{equation}
and at the solid substrate the impermeability condition (\ref{bcp1})
\begin{equation}
\left.u_z\right|_{z=0}=\left.\partial_z \phi\right|_{z=0}=0,\quad 1<r<\infty. \label{bc1}
\end{equation}
We express the (dimensionless) pressure in the liquid using the linearized Bernoulli integral
\begin{equation}
p=\frac{a p_0}{\sigma}-\zeta \phi,
\end{equation}
and, neglecting the viscous stress, find for the dynamic boundary condition at the gas-liquid interface (\ref{bc4})
\begin{equation}
PH-\partial_r^2\eta-\frac{1}{r}\partial_r\eta=\zeta \left.\phi\right|_{z=0},\quad 0\leq r<1,\label{dyncondpot}
\end{equation}
with 
\begin{equation}
P=\frac{a\kappa\lambda p_0}{\sigma}, \label{presp}
\end{equation}
the ratio of the gas stiffness to the surface tension stiffness.

\subsection{Reduction to an eigenvalue problem}
As shown in Appendix \ref{appf}, the Hankel transform can be used to reduce (\ref{bc2})-(\ref{dyncondpot}) to a set of dual integral equations in terms of $\phi$.
The solution of this system results in the following expression for the dynamic boundary condition (\ref{dyncondpot}) 
\begin{equation}
PH -\partial_r^2\eta-\frac{1}{r}\partial_r\eta =-  \zeta^2 \int_0^\infty 
v(k) J_0(kr)\mathrm{d}k,\label{dyncond2}
\end{equation}
with
\begin{equation}
 v(k)= \int_0^1 \eta(s) s J_0(ks)\mathrm{d}s. \label{uk}
\end{equation}
Integration of (\ref{dyncond2}) leads to the following eigenvalue problem to be solved for $\zeta$ and $\eta$
\begin{equation}
\eta(r)=\frac{1}{4}PH(r^2-1)+ \zeta^2 \int_0^\infty 
\frac{v(k)}{k^2}\left[J_0(k)- J_0(kr)\right]\mathrm{d}k\label{dyncond2b}.
\end{equation}
To find the solution of this integral equation we expand the interface deformation $\eta$ into a Fourier-Bessel series
\begin{equation}
\eta(r,t)=\sum_{k=1}^\infty c_k(t) J_0(j_kr),\label{fbes}
\end{equation}
with $j_k$ denoting the $k$th zero of the Bessel function $J_0$. 
Substituting (\ref{fbes}) into (\ref{dyncond2b}) and taking the inner product with $rJ_0(j_n r)$, we obtain the following generalized eigenvalue problem for the eigenfrequency $\zeta$ (see Appendix \ref{appf} for details) 
\begin{eqnarray}
2P \sum_{k=1}^\infty\frac{ J_1(j_k)}{j_k} \frac{J_1(j_n)}{j_n}c_k+\frac{1}{2}c_ nj_n^2 J_1^2(j_n)=- \zeta^2\sum_{k=1}^\infty c_kJ_1(j_k)  J_1(j_n)j_kj_n f(j_k,j_n),
\label{eqcoef}
\end{eqnarray}
with $f(j_k,j_n)$ given by (\ref{fjk}). 

\section{Weak viscous effects}\label{cpotflow}
In a real flow, viscous dissipation in both the bulk of the liquid and the boundary layer on the solid surface dampen the bubble oscillations. The ratio of dissipation in the boundary layer to dissipation in the bulk is given by $a/\delta$~\cite{Batchelor}, where $\delta\sim\sqrt{\nu/\omega}$ is the viscous boundary layer thickness. If we scale the angular frequency on the basis of the free bubble Minnaert frequency, $\omega\sim \left(\kappa p_0/\rho\right)^{1/2}/a$, we find that the ratio of damping in the boundary layer to damping in the bulk is given by
\begin{equation}
\frac{\kappa p_0a^2}{\rho\nu^2} =\frac{P}{\lambda \mathrm{Oh}^2},
\end{equation}
with Oh the Ohnesorge number, defined by
\begin{equation}
\mathrm{Oh}=\sqrt{\frac{\rho \nu^2}{\sigma a}}.\label{ohne}
\end{equation}
This dimensionless parameter is a measure of the damping during one period of oscillation.
In case the bulk dissipation dominates, i.e. for smaller pits, we can use the potential flow solution to estimate the damping coefficient~\cite{Batchelor}.

To describe the oscillations of the damped system, we use a Lagrangian formulation complemented by the Rayleigh dissipation function. 
We again express the interface shape in terms of the Fourier-Bessel series (\ref{fbes}).
The motion of the system is now given by
\begin{equation}
\frac{\partial}{\partial t}\left(\frac{\partial \mathcal{L}}{\partial \dot{c_k}} \right)-\frac{\partial \mathcal{L}}{\partial c_k}=-\mathrm{Oh}\frac{\partial \mathcal{R}}{\partial \dot{c}_k},\label{disseq}
\end{equation}
with $\dot{c}_k=\zeta c_k$, $\mathcal{L}=\mathcal{E}_k-\mathcal{E}_p$ the Lagrangian, $\mathcal{R}$ the Rayleigh dissipation function,  defined as $\mathcal{R}=\mathcal{D}/2$ with $\mathcal{D}$ the rate of viscous dissipation in the liquid~\cite{Landau}.
To find an expression for $\mathcal{L}$ in terms of $c_k$, we calculate the kinetic energy $\mathcal{E}_k$ and potential energy $\mathcal{E}_p$ of the system. The dimensionless kinetic energy of the liquid can be expressed as~\cite{Lamb} 
\begin{equation}
\mathcal{E}_k=\frac{1}{2}\int_V u_i^2 \mathrm{d}V=\frac{1}{2}\int_A \phi \frac{\partial \phi}{\partial x_i}n_i \mathrm{d}A,\label{ekin}
\end{equation}
where $\boldsymbol{n}$ represents the unit surface normal directed out of the liquid. Due to the impermeability condition (\ref{bc1}), the integral (\ref{ekin}) reduces to an integral over the bubble interface:
\begin{equation}
\mathcal{E}_k=-\pi\int_0^1 \phi(r,0)\partial_t\eta~r\mathrm{d}r.\label{ekin2}
\end{equation}
The potential energy of the system can increase by an increase in area through the effect of surface tension or by a decrease in volume through compression of the gas
\begin{equation}
\mathcal{E}_p= \int_{A_0}^A\mathrm{d}A-\int_{V_0}^V(p_g-p_0)\mathrm{d}V=\pi \int_0^1 \left(\partial_r\eta\right)^2r\mathrm{d}r+\frac{1}{2} \pi P H^2,\label{ep2}
\end{equation}
with $P$ given by (\ref{presp}).
The rate of viscous dissipation in potential flow reads~\cite{Lamb}
\begin{equation}
\mathcal{D}=\frac{1}{2}\mathrm{Oh}\int_V \left(\frac{\partial u_i}{\partial x_k}+\frac{\partial u_k}{\partial x_i}\right)^2\mathrm{d}V=2\mathrm{Oh}\int_A u_i\frac{\partial u_i}{\partial x_k}n_k\mathrm{d}A.\label{dis}
\end{equation}
By integrating (\ref{dis}) over the surface we arrive at
\begin{equation}
\mathcal{D}=-8\pi\mathrm{Oh}\int_0^1 \left.\frac{\partial\phi}{\partial r}\right|_{z=0}\partial_r\eta ~r\mathrm{d}r.\label{dis2}
\end{equation}
Substituting (\ref{fbes}) into (\ref{ekin2}), (\ref{ep2}), and (\ref{dis2}), we can express the kinetic energy, potential energy, and dissipation in terms of the degrees of freedom $c_k$ and $\dot{c}_k$ as
\begin{eqnarray}
\mathcal{E}_k&=&\pi \sum_k^\infty\sum_l^\infty\dot{c}_k\dot{c}_nj_kj_nJ_1(j_k)J_1(j_n)f(j_k,j_n),\label{ekinnum}\\
\mathcal{E}_p&=&\frac{1}{2}\pi\sum_{k=1}^\infty j_k^2c_k^2J_k^2(j_k)
+2\pi P \sum_{k=1}^\infty\sum_{l=1}^\infty\frac{c_k}{j_k}\frac{c_l}{j_l}J_1(j_k)J_1(j_l),\label{epnum}\\
\mathcal{D}&=&8\pi\mathrm{Oh}\sum_k^\infty\sum_l^\infty\dot{c}_k\dot{c}_nj_kj_nJ_1(j_k)J_1(j_n)g(j_k,j_n).\label{disnum}
\end{eqnarray}
with $f(j_k,j_n)$ given by (\ref{fjk}), and $g(j_k,j_n)$ given by (\ref{gjk}).
Substituting (\ref{ekinnum})-(\ref{disnum}) into (\ref{disseq}) and replacing $\dot{c}_k$ by $\zeta c_k$, we obtain
\begin{eqnarray}
2P \sum_{k=1}^\infty \frac{ J_1(j_k)}{j_k} \frac{J_1(j_n)}{j_n}c_k+\frac{1}{2}c_ nj_n^2 J_1^2(j_n)=- \zeta^2\sum_{k=1}^K c_kJ_1(j_k)  J_1(j_n)j_kj_nf(j_k,j_n)\nonumber\\ -4\zeta\mathrm{Oh}\sum_{k=1}^K c_k J_1(j_n)J_1(j_k)j_nj_k g(j_n,j_k).\label{diseqnum}
\end{eqnarray}
Note that putting Oh to zero in (\ref{diseqnum}), i.e. neglecting the viscous dissipation, leads to exactly the same equation as derived before for potential flow, which reconfirms (\ref{eqcoef}).

\section{Unsteady Stokes flow}\label{stokesflow}
In the previous sections potential flow was used to calculate the pressure and velocities in the liquid. On the solid substrate, however, the no-slip boundary condition (\ref{bcp3a}) applies. Hence, a viscous boundary layer develops on the substrate, which cannot be accounted for in a potential flow model. Therefore, we repeat the calculation of the liquid pressure using an unsteady Stokes flow model. For small interface deformations $\eta\ll a$, as is the case here, the nonlinear term of the Navier Stokes equations can be neglected with respect to the unsteady inertia term, and the unsteady Stokes equations describe the flow in the entire domain~\cite{Batchelor}.
\subsection{Governing equations}
The unsteady Stokes equations in dimensionless form, with the dimensionless quantities as defined in (\ref{dimpar}), read
\begin{equation}
\zeta \boldsymbol{u}=-\nabla p+\mathrm{Oh} \nabla^2\boldsymbol{u},\label{stokes}
\end{equation}
with Oh the Ohnesorge number defined in (\ref{ohne}). We express the velocity in terms of a stream function $\Psi$ as 
\begin{equation}
\boldsymbol{u}=\nabla \times \left(\frac{1}{r}\Psi \boldsymbol{e}_\theta\right)=-\frac{\partial }{\partial z}\left(\frac{1}{r}\Psi\right)\boldsymbol{e}_r+\frac{1}{r}\frac{\partial \Psi}{\partial r} \boldsymbol{e}_z.\label{ustream}
\end{equation}
Taking the curl of (\ref{stokes}), we obtain
\begin{equation}
\zeta \boldsymbol{\Omega}=\mathrm{Oh} \nabla^2 \boldsymbol{\Omega},\label{omeq}
\end{equation}
with vorticity $\boldsymbol{\Omega}=\nabla\times \boldsymbol{u}=-\nabla^2 \left(\Psi/r  \boldsymbol{e}_\theta\right)$. 
As boundary conditions we have again the kinematic condition (\ref{bcp2}) and impermeability of the substrate (\ref{bcp1}). As mentioned before, we impose a no-slip condition on the solid (\ref{bcp3a}) as well as on the bubble surface (\ref{bc3}), to render the mathematical problem tractable.
The dynamic boundary condition now reads
\begin{equation}
PH-\partial_r^2\eta-\frac{1}{r}\partial_r\eta=\left.p\right|_{z=0},\quad 0\leq r<1,\label{dyncondstok}
\end{equation}
with the pressure to be calculated from (\ref{stokes}).
Note that the normal viscous stress drops out from (\ref{dyncondstok}) as a consequence of (\ref{bc3}) which, from the equation of continuity, implies that $\partial u_z/\partial z=0$ on $z=0$.

\subsection{Reduction to an eigenvalue problem}
The solution to (\ref{omeq}) can again be expressed in terms of the Hankel transform.
Then, the eigenvalue problem to be solved for $\zeta$, $\eta$ is given by (details of the calculation can be found in Appendix \ref{apps})
\begin{equation}
\eta(r)=\frac{1}{4}PH(r^2-1)- \int_0^\infty 
\frac{v(k)}{k^2}\left[J_0(k)- J_0(kr)\right]\left[\zeta^2+\zeta \mathrm{Oh} \left( k^2+k\sqrt{k^2+\zeta/\mathrm{Oh}}\right) \right]\mathrm{d}k
,\quad 0\leq r<1.\label{evpstokes1}
\end{equation}
Again, we expand the interface deformation $\eta$ into the Fourier-Bessel series (\ref{fbes}), and obtain (see Appendix \ref{apps} for details)
\begin{eqnarray}
& &2 P\sum_{k=1}^\infty \frac{c_k}{j_k j_n} J_1(j_k) J_1(j_n) +\frac{1}{2} c_n j_n^2 J_1^2(j_n) =\nonumber \\
& &\sum_{k=1}^\infty c_k j_k j_n J_1(j_k) J_1(j_n) \int_{0}^{\infty} \frac{ J_0^2(s ) }{(j_k^2 -s^2) (j_n^2 -s^2)} \left(\zeta^2+\zeta \mathrm{Oh}\left[s^2+s \sqrt{s^{2}+\zeta/\mathrm{Oh}}\right]\right)   \mathrm{d}s.\label{evpstokes}
\end{eqnarray}

\section{Numerical solution method}
To find the resonance frequency and interface shape in the potential flow model, the generalized eigenvalue problem (\ref{eqcoef}) is truncated to $K$ terms and solved numerically with Mathematica 8 (Wolfram Research) for $\zeta$ and $c_k$; see Appendix \ref{appf}. We studied the convergence of the sum (\ref{fbes}) for the first three modes by taking up to $K=100$ terms into account for $P=0$ to $P=500$. We found that the system converges rapidly: for the lowest mode $K=3$ was already sufficient for accurate reconstruction of the interface shape. For higher modes, the matrix size increases because more Bessel functions are required to describe the interface shape: for mode 3, we used $K=6$. The larger the matrix, the more eigenfrequencies can be calculated. 

The generalized eigenvalue problem (\ref{evpstokes}) for the Stokes flow model has to be solved iteratively, due to the complexity of the integral. To this end, we split the integral into three parts, so that the equation to be solved becomes
\begin{eqnarray}
& &2 P\sum_{k=K}^\infty \frac{c_{k}}{j_k j_n} J_1(j_k) J_1(j_n) +\frac{1}{2} c_{n} j_n^2 J_1^2(j_n) =\nonumber \\
& &\sum_{k=1}^\infty c_{k} j_k j_n J_1(j_k) J_1(j_n)\left[\zeta^2 f(j_k,j_n)+\zeta \mathrm{Oh}\left\{ g(j_k,j_n)+h(j_k,j_n, \zeta, \mathrm{Oh})\right\}\right],\label{eigvstokesht}
\end{eqnarray}
with $f(j_k,j_n)$ given by (\ref{fjk}), $g(j_k,j_n)$ by (\ref{gjk}), and $h(j_k,j_n)$ by (\ref{inth}).
To reduce (\ref{systs}) to a generalized eigenvalue problem that can be solved with Mathematica, we write
$\mathbf{d}=\zeta \mathbf{c}$, so that the resulting system becomes
\begin{equation}
\begin{pmatrix} \mathsf{\emptyset} & \mathsf{I} \\ 2P\mathsf{B}+\frac{1}{2}\mathsf{C}\quad & \mathrm{Oh}\left[\mathsf{G}+\mathsf{H(\zeta^{(i)},\mathrm{Oh})}\right] \end{pmatrix} \left( \begin{array}{c} \mathbf{c}^{(i+1)} \\ \mathbf{d}^{(i+1)} \end{array} \right)=\zeta^{(i+1)}\begin{pmatrix} \mathsf{I} & \mathsf{\emptyset} \\ \mathsf{\emptyset} & -\mathsf{A} \end{pmatrix} \left( \begin{array}{c} \mathbf{c}^{(i+1)} \\ \mathbf{d}^{(i+1)} \end{array} \right),\label{systsr}
\end{equation}
with $i$ the iteration number, $\mathbf{c}$ a vector with elements $c_k$, $\mathsf{\emptyset}$ the zero matrix , $\mathsf{I}$ the unit matrix, $\mathsf{A}$ given by (\ref{acoef}), $\mathsf{B}$ by (\ref{bcoef}), $\mathsf{C}$ by (\ref{ccoef}),
$\mathsf{G}$ by (\ref{gcoef}), and $\mathsf{H}$ by (\ref{hcoef}). The matrices $\mathsf{A}$, $\mathsf{G}$, and $\mathsf{H}$ correspond to the three terms in which the integral in (\ref{evpstokes}) is decomposed.
The generalized eigenvalue problem (\ref{systsr}) is solved iteratively with the modified potential-flow solution used as initial guess $\zeta^{(1)}$ in matrix $\mathsf{H}$.
With this initial guess, we evaluate the integral (\ref{inth}) numerically, which we then use to solve (\ref{systsr}). This step permits an improved estimate $\zeta^{(2)}$ of the eigenvalue, which is substituted again into the matrix $\mathsf{H}$. This procedure is repeated until convergence is reached, i.e. until the difference in both frequency and damping between the current and the previous iteration is less than $0.01\%$ of the current result.
Again, we investigated the influence of the matrix size on the calculation of the eigenfrequency and interface shape for the lowest three modes by taking a system size up to $K=16$.
For the lowest two modes, $K=3$ was sufficiently accurate to calculate the eigenfrequency, whereas for mode 3, $K=6$ was used. Convergence of (\ref{systsr}) was achieved after 4 iterations for the three lowest modes; see Appendix \ref{apps} for details. To further check the convergence of the solutions obtained, we used the potential flow solution as initial guess, and slowly increased the Ohnesorge number from 0 to Oh$_s=0.0303$. Using this method, we obtained the same results for the resonance frequency and the damping as by starting directly at Oh$_s$ with the modified potential flow solution as initial guess.

\section{Results}
\begin{figure}
\begin{center}
	\includegraphics[width=7 in]{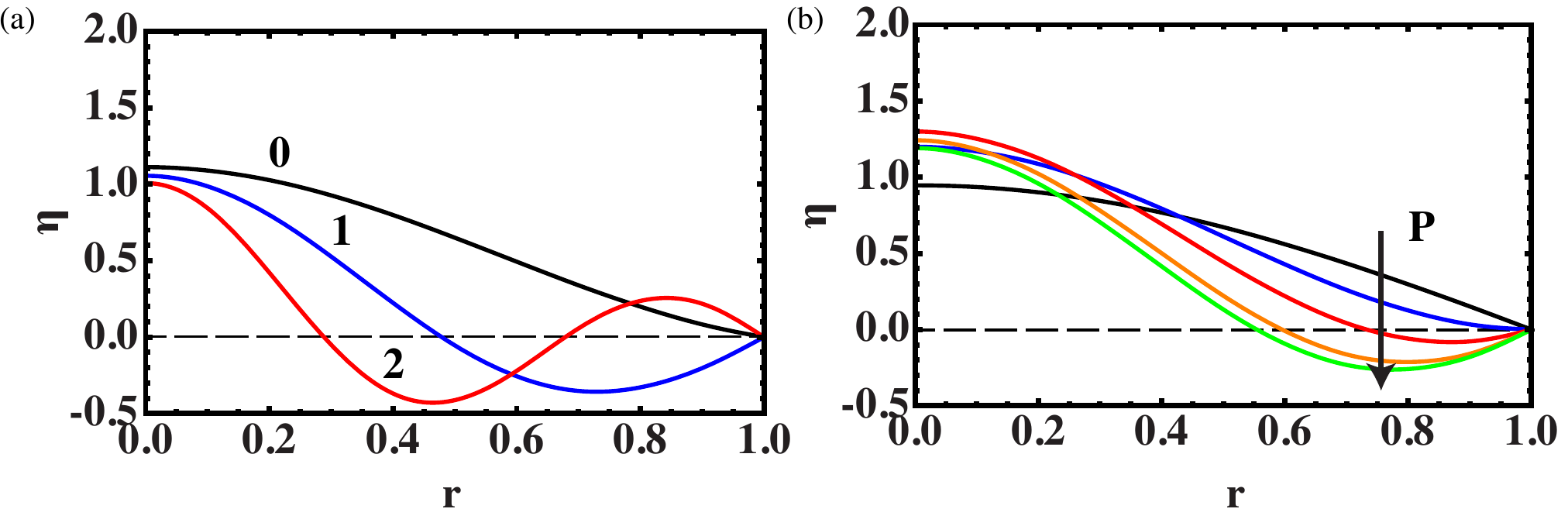}
\caption{(Color online) The interface shape as calculated from the potential flow model for (a) $P=20.5$, mode 0  (black), mode 1 (blue online), and mode 2 (red online). (b) The interface shape for mode 0 with $P=0$ (black), 40 (blue online), 80 (red online), 160 (orange online), 320 (green online). Note that for $P\ge 80$, mode 0 has an extra node in addition to the one at the rim of the pit.}\label{shapes}
\end{center}
\end{figure}

In the generalized eigenvalue problem for $\zeta$, $\eta$, only two dimensionless parameters appear: the ratio of the gas stiffness to the surface tension stiffness, $P$ (\ref{presp}), and the Ohnesorge number Oh (\ref{ohne}). For a gas pocket with $\lam=1$ in a $15$-$\mu$m cylindrical micropit submerged in water under standard conditions, the corresponding values of the dimensionless groups are $P=P_s=20.5$ and Oh=Oh$_s$=0.0303. In this example, we find the dimensional resonance frequency for the first three modes in potential flow to be 121, 274, and 556 kHz, respectively. The first result is not very different from the frequency estimate (\ref{miller}) by Miller \& Nyborg~\cite{Miller:1983}  which is 151 kHz for the lowest mode of a 15-$\mu$m pit. The interface shape for the first three modes is depicted in Figure \ref{shapes}a. One can see that, with these parameter values, the largest contribution to the interface shape of mode 0 comes from the first term in the Fourier-Bessel series (\ref{fbes}), whereas for mode 1 the largest contribution comes from the second term, etc.

\begin{figure}
	\includegraphics[width=6.5 in]{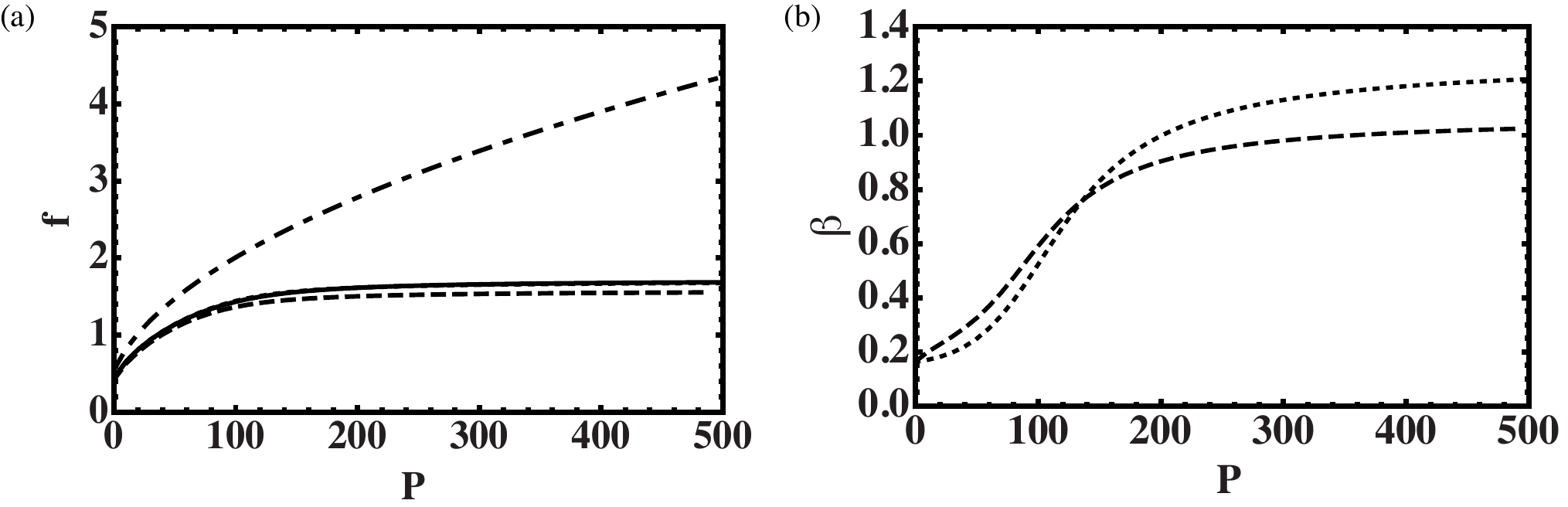}
\caption{(a) The dimensionless frequency $f=\omega/2\pi$ of mode 0 versus the dimensionless parameter $P$, defined in (\ref{presp}), for potential flow (black, solid), modified potential flow (black, dotted) and Stokes flow (black, dashed) with Oh=0.0303. The results for potential flow and modified potential flow nearly overlap. For comparison, the result of Miller \& Nyborg\cite{Miller:1983} (\ref{miller}) is also shown (dash-dotted). (b) Dimensionless damping $\beta$ versus $P$ for the modified potential flow and Stokes flow models. }\label{parstudPm0}
\end{figure}

\begin{figure}
	\includegraphics[width=7 in]{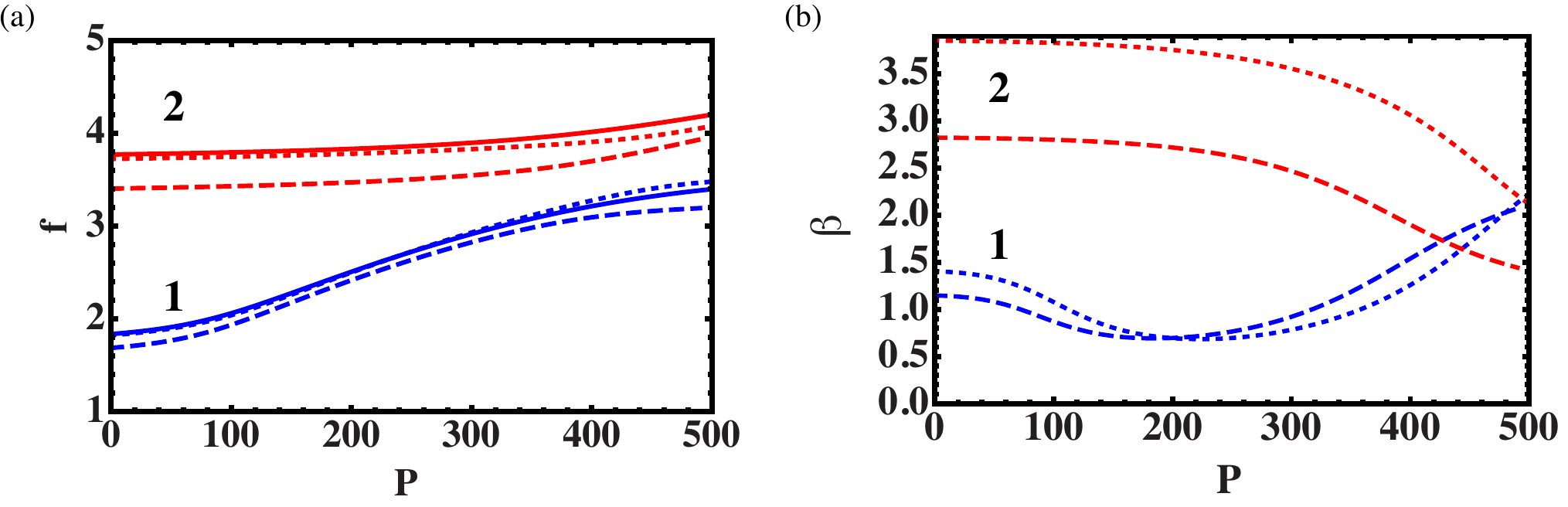}
\caption{(Color online) (a) The dimensionless frequency $f$ of mode 1 (blue online), and mode 2 (red online), versus the dimensionless parameter $P$, defined in (\ref{presp}), for potential flow (solid), modified potential flow (dotted) and Stokes flow (dashed) with Oh=0.0303. A plateau in the frequencies similar to that for mode 0 is observed, but it occurs at larger $P$ (not shown in the figure for mode 2). (b) Dimensionless damping $\beta$ versus $P$ for the different cases. }\label{parstudP}
\end{figure}

\begin{figure}
	\includegraphics[width=7 in]{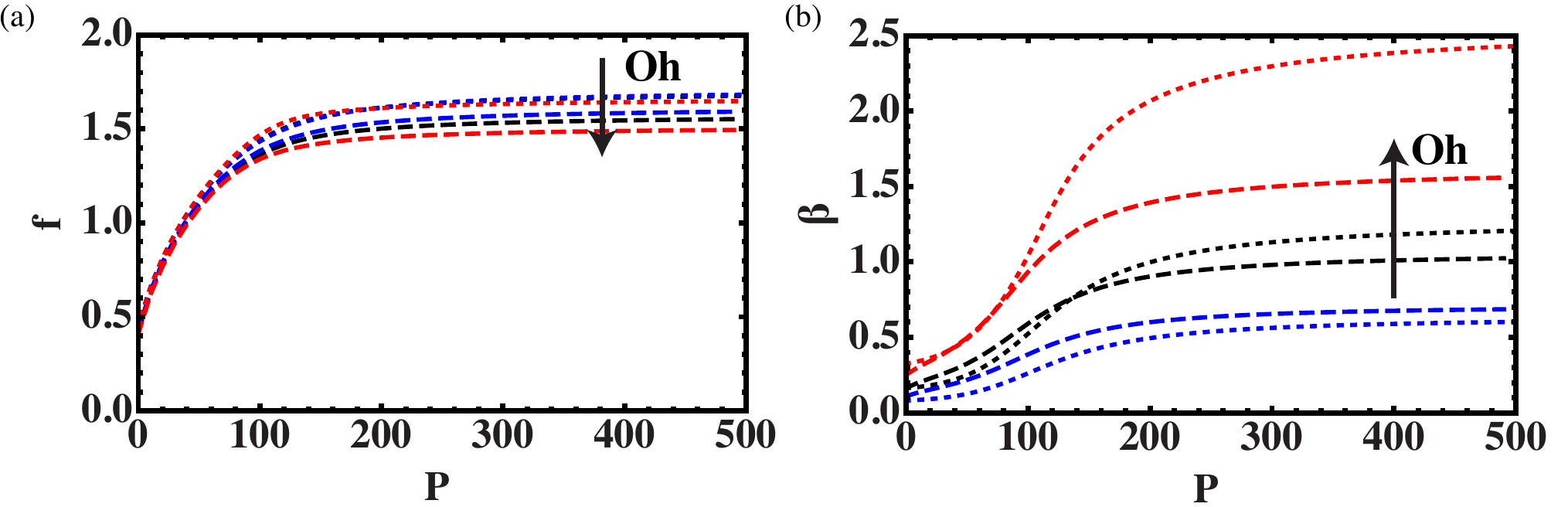}
\caption{(Color online) (a) The dimensionless frequency $f$ of mode 0 versus the dimensionless parameter $P$, defined in (\ref{presp}), for modified potential flow (dotted), and Stokes flow (dashed) with Oh=0.01515 (blue online), Oh=0.0303 (black), and Oh=0.0606 (red online). (b) Dimensionless damping $\beta$ versus $P$ for the different cases.}\label{parstudoh}
\end{figure}

Figures \ref{parstudPm0}, \ref{parstudP}, and \ref{parstudoh} show how the resonance frequency and damping coefficient depend on the two dimensionless parameters $P$ and Oh. As expected, both the damping and frequency increase with the mode number 
with the result that, after a generic initial perturbation, the bubble will oscillate the longest at its fundamental resonance frequency whereas 
higher frequencies dampen out earlier. The difference in resonance frequency between the potential flow (PF), modified potential flow (mPF), and Stokes flow (SF) models is very small which means that, in the parameter range of interest, the resonance frequency is mainly determined by inertia and can be obtained from the potential flow model with sufficient accuracy.

\begin{figure}
\begin{center}
	\includegraphics[width=3.5 in]{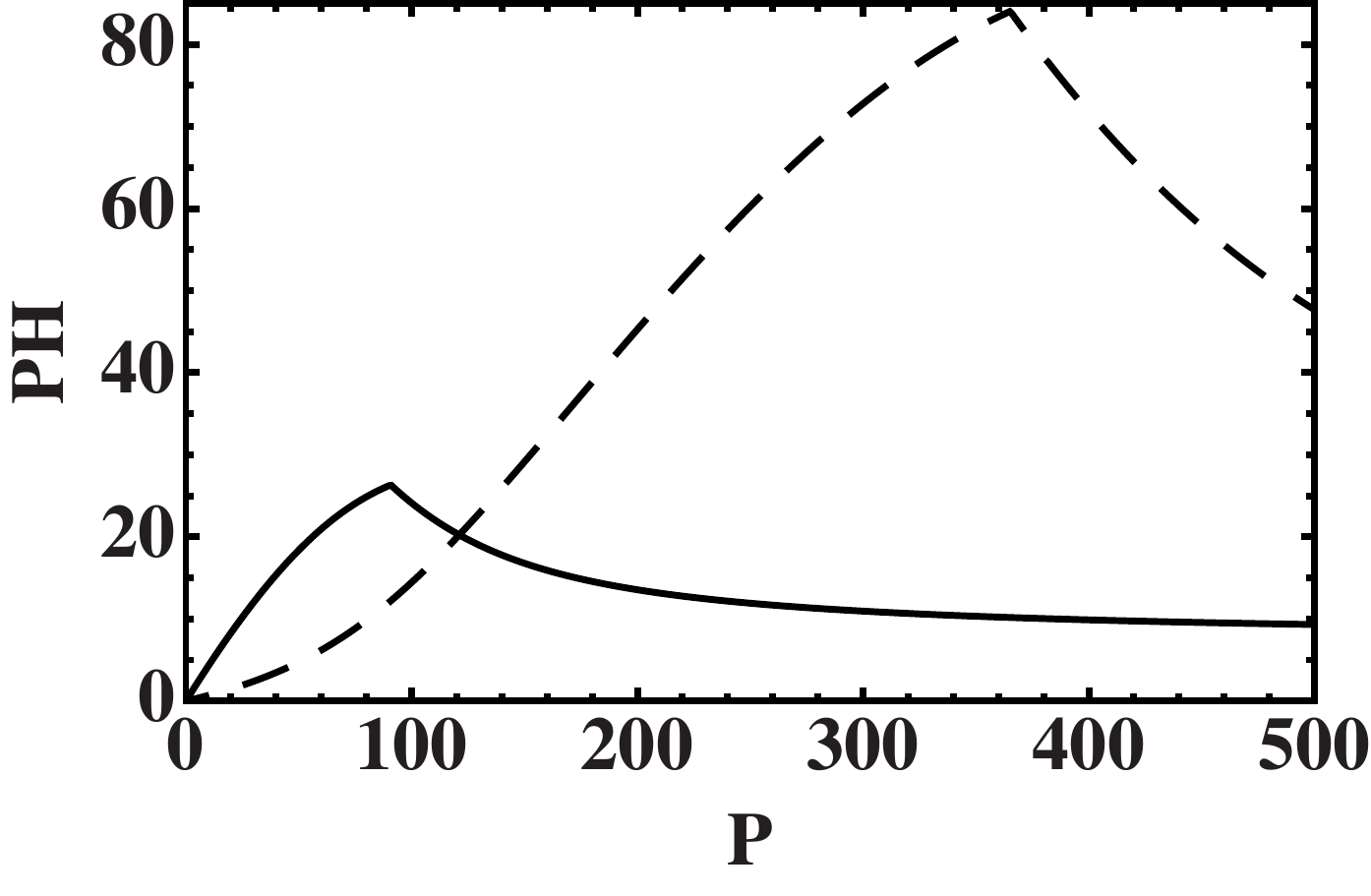}
\caption{(Color online) The relative stiffness of the gas pocket times its volume change, $PH$, versus $P$, for mode 0 (solid) and mode 1 (dashed). Initially $PH$ increases, until a maximum is reached, then it tends to a finite value.}\label{ph}
\end{center}
\end{figure}

\begin{figure}
\begin{center}
	\includegraphics[width=7 in]{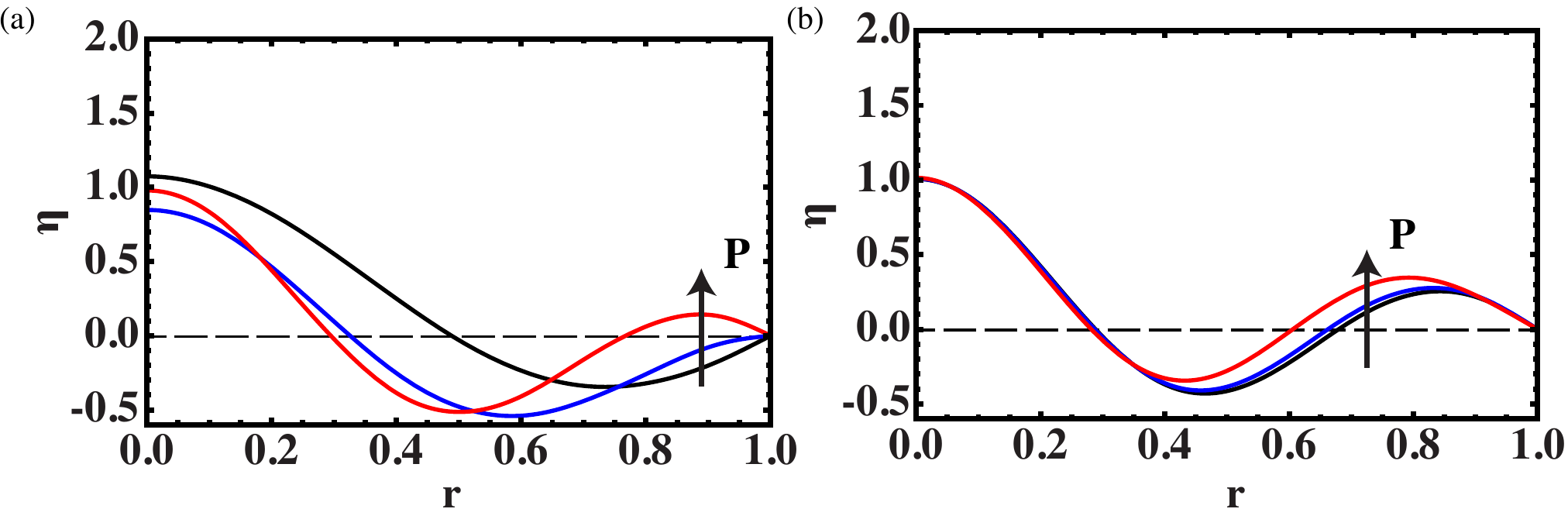}
\caption{(Color online) The interface shape as calculated from the potential flow model with Oh=0.0303 for $P=0$ (black), $P=250$ (blue online), and $P=500$ (red online) for (a) mode 1 and (b) mode 2. Note that, for $P=500$, mode 1 has developed an extra node.}\label{shapes2}
\end{center}
\end{figure}

In Fig. \ref{parstudPm0}a, the graph of the frequency $f=\omega/2\pi$ versus $P$ for mode 0 shows that the resonance frequency first increases with $P$, until it levels off to a dimensionless value $f=1.72$ for $P>200$, approximately. In the approximate solution (\ref{miller}) by Miller\cite{Miller:1983} such a plateau is not observed. Figure \ref{parstudP}a shows graphs of $f$ versus $P$ for modes 1 and 2. Here, a similar increase in $f$ with $P$ is observed, but the plateau is reached at larger values of $P$.
Initially, the resonance frequency increases with $P$ because at larger $P$ it becomes more difficult to change the volume of the gas, and hence the system becomes stiffer, which leads to a higher resonance frequency. The reason for the occurrence of a plateau in the frequency lies in the increasing stiffness of the gas.  As $P$ increases and the system becomes stiffer, it becomes more difficult to decrease the gas volume change $H$, as defined in (\ref{eqH}), and the system responds by increasing the area of the interface instead (see Fig. \ref{shapes}a).  To further decrease the volume change, at some point an extra node has to appear in the interface shape, as can be seen in Fig. \ref{shapes}a for $P\ge 80$.
This node is pushed towards the axis of the pit as $P$ is increased further. In this way, the interface area increases more and more, and the net volume change due to the surface elevation eventually tends to zero, which means that $V\to V_0$. Figure \ref{ph} shows that this decrease in the amplitude of the volume oscillations occurs in such a way that the product $PH$ tends to a constant value. Hence, the resonance frequency levels off, and the system eventually oscillates with a fixed interface shape. 
The interface shapes corresponding to modes 1 and 2 are depicted in Fig. \ref{shapes2}. For mode 1, the extra node appears around $P\simeq 500$, whereas for mode 2, it will occur at a larger $P$.


The parameter $P$ also has an effect on the damping coefficient as shown in Figs. \ref{parstudPm0}b and \ref{parstudP}b. For mode 0 the damping coefficient increases with $P$, until a final plateau is reached. For mode 1, however, a minimum is observed and the plateau is reached for larger $P$. For mode 2, the damping decreases and reaches a minimum beyond the maximum value of $P$ shown in the graph. The presence of a minimum could be explained as follows: as the stiffness of the gas increases with $P$, the relative volume change decreases, which leads to a smaller liquid displacement and viscous energy dissipation. However, as the frequency increases, the damping increases as well. These two effects compete, and give rise to a minimum in the damping coefficient. The plateau is reached at larger $P$, when the product $PH$ tends to a finite value (see also Fig. \ref{ph}). The difference in damping coefficients between the mPF and SF models is larger for the higher modes. One also observes that, for some values of $P$, the mPF damping is even larger than the SF damping. This behavior is due to the difference in velocity profiles between the mPF and SF, and is also known to occur for shape oscillations  of drops and bubbles~\cite{Prosperetti:1980}.

Whereas $P$ has a large influence on the resonance frequency of the pit, the influence of Oh is only very small as shown in Fig. \ref{parstudoh}. The influence of the Ohnesorge number on the damping coefficient is of course large.  

\section{Conclusion}
The resonance frequency, damping and interface shape of a gas pocket entrapped on the surface of an submerged solid have been calculated. To describe the hydrodynamic problem in the liquid domain, both a potential and an unsteady Stokes flow model have been used. The potential flow model gives a reliable prediction of the resonance frequency of the gas pocket, which is mainly determined by inertia in the liquid.
To derive an estimate for the damping of the oscillations, the bulk dissipation was calculated from the potential flow model. A more accurate prediction of the damping was derived based on the unsteady Stokes flow model, which is valid throughout entire domain and therefore includes the contributions of both the boundary layer and the bulk. However, the Stokes flow results will overestimate the real damping somewhat in the case that the liquid-gas interface is clean: in the method described here, a no-slip condition on the free surface was used, which leads to some additional dissipation. 

The resonance frequency, damping and interface shape of an entrapped gas pocket depend on two dimensionless numbers: the ratio $P$ of the gas stiffness to the surface tension stiffness, defined in (\ref{presp}), and the Ohnesorge number (\ref{ohne}), which represents the relative importance of viscous forces. In general, the resonance frequency increases with increasing gas stiffness.
However, an unexpected feature of our results is that, when the volume stiffness of the gas pocket greatly exceeds the surface stiffness, the normal modes develop an extra node, and the resonance frequency tends to an asymptotic value.

\begin{acknowledgements}
We would like to thank Laura Stricker, Jacco H. Snoeijer and Detlef Lohse for valuable discussions.
We acknowledge the financial support of the NWO-Spinoza program.
\end{acknowledgements}

\appendix
\section{Potential flow solution}\label{appf}
To obtain a solution for the velocity potential $\phi$ we set
\be
  \pat_z \phi=\int_0^\infty k \Phi(k) J_0(kr)e^{-kz} \mathrm{d}k
\ee
and, upon integration, find that
\be
 \phi =-\int_0^\infty \Phi(k) J_0(kr)e^{-kz} \mathrm{d}k.
\label{exppot}
\ee
Using the boundary conditions (\ref{bc2}) and (\ref{bc1}) and the orthogonality relation for Bessel functions we obtain 
\be
\int_0^\infty \int_0^\infty kr \Phi(k) J_0(kr) J_0(hr) \mathrm{d}r\mathrm{d}k =\int_0^\infty \Phi(k) \delta(h-k)\mathrm{d}k =
\zeta\int_0^1 \eta(r) r J_0(hr)\mathrm{d}r ,
\ee
and, therefore, 
\be
 \Phi(h) =\zeta\int_0^1 \eta(r) r J_0(hr)\mathrm{d}r=\zeta v(h). \label{coefP}
\ee
Thus
\be
 \phi(r,0) =-\zeta\int_0^\infty  v(k) J_0(kr)\mathrm{d}k.\label{phi0}
\ee
Substituting this result into (\ref{dyncondpot}) we find (\ref{dyncond2}).

The next step is to express the interface deformation $\eta$ in terms of the Fourier-Bessel series (\ref{fbes}), to obtain
\bea
v(s)=\sum_{k=1}^\infty c_k \int_0^1  r\, J_0(sr)J_0(j_kr)\mathrm{d}r=
\sum_{k=1}^\infty c_k \frac{j_kJ_0(s)J_1(j_k)}{j_k^2-s^2},\label{uk2}
\eea
\be
 H=2 \sum_{k=1}^\infty c_k \int_0^1 r  J_0(j_kr)\mathrm{d}r  = 2 \sum_{k=1}^\infty \frac{c_k}{j_k} J_1(j_k).
\label{hetaser1}
\ee
Substitution into (\ref{dyncond2b}) leads to
\be
\sum_{k=1}^K c_kJ_0(j_k r)= \frac{1}{2}\mathrm{P} \sum_{k=1}^\infty \frac{c_k}{j_k} J_1(j_k) (r^2-1)
+\zeta^2\sum_{k=1}^\infty c_kj_kJ_1(j_k) 
\int_0^\infty \frac{\left[J_0(s)-J_0(sr)\right]J_0(s)}{(j_k^2-s^2)s^2} \mathrm{d}s.\label{eqcoef2}
\ee
To obtain an equation for each of the unknowns, we multiply (\ref{eqcoef2}) by $rJ_0(j_nr)$ and integrate between 0 and 1, to find
\bea
&&2P \sum_{k=1}^\infty\frac{c_k}{j_kj_n} J_1(j_k)J_1(j_n)
+\frac{1}{2}c_ nj_n^2 J_1^2(j_n) = 
-\zeta^2 \sum_{k=1}^\infty c_kj_kJ_1(j_k) 
\int_0^\infty \frac{J_0(s)}{j_k^2-s^2} \mathrm{d}s\int_0^1 rJ_0(sr)J_0(j_nr)\mathrm{d}r
\nonumber \\ && =
-\zeta^2\sum_{k=1}^\infty c_kJ_1(j_k)j_kj_n J_1(j_n)f(j_k,j_n),
\eea
with 
\bea
f(j_k,j_n)&=&\int_0^\infty \frac{J_0^2(s)}{(s^2-j_k^2)(s^2-j_n^2)}ds\nonumber\\
&=&
\left\{ {\begin{array}{ccc}
\frac{4}{\pi}\frac{1}{j_k^2-j_n^2}\left[{_2F_3}(1,1;\tfrac{3}{2},\tfrac{3}{2},\tfrac{3}{2};-j_n^2)-{_2F_3}(1,1;\tfrac{3}{2},\tfrac{3}{2},\tfrac{3}{2};-j_k^2)\right],&\mathrm{if}& k\neq n,\\
\frac{32}{27\pi}{_2F_3}(2,2;\tfrac{5}{2},\tfrac{5}{2},\tfrac{5}{2};-j_k^2),&\mathrm{if}& k= n,\label{fjk}
\end{array} } \right.
\eea
where $_2F_3$ is the hypergeometric function~\cite{Abramowitz}. 
After truncation of the Fourier-Bessel series (\ref{fbes}) to $K$ terms, we can express the eigenvalue problem in matrix form as
\begin{equation}
\left(\zeta^2\mathsf{A}+2P\mathsf{B}+\frac{1}{2}\mathsf{C}\right)\mathbf{c}=\mathbf{0},\label{systpot}
\end{equation}
with $\mathsf{A}$ a $K\times K$-matrix with coefficients 
\begin{equation}
A_{kn}=J_1(j_k)  J_1(j_n)j_kj_n f(j_k,j_n),\label{acoef}
\end{equation}
$\mathsf{B}$ a $K\times K$-matrix with coefficients
\begin{equation}
B_{kn}=\frac{J_1(j_k)J_1(j_n)}{j_kj_n},\label{bcoef}
\end{equation}
and $\mathsf{C}$ a $K\times K$ diagonal matrix with coefficients
\begin{equation}
C_{kn}=j_k^2J_1^2(j_k)\delta_{kn}, \label{ccoef}
\end{equation}
and $\mathbf{c}$ a $K$-array with coefficients $c_k$.
The generalized eigenvalue problem (\ref{systpot}) is the solved with Mathematica 8 (Wolfram Research).

\section{Stokes flow solution}\label{apps}
The general solution to (\ref{omeq}) in terms of the Hankel transform reads
\begin{equation}
\Omega(r,z)=\int_{0}^{\infty}G(k)J_{1}(kr)e^{-z\sqrt{k^{2}+\zeta/\mathrm{Oh}}}\mathrm{d}k, \label{omsol}
\end{equation}
with $G$  to be determined from the boundary conditions (\ref{bcp2}), (\ref{bcp1}), (\ref{bcp3a}), and (\ref{bc3}). From (\ref{omsol}) we can determine an expression for stream function $\Psi$ using
\begin{equation}
\boldsymbol{\Omega}=-\nabla^2 \left(\frac{\Psi}{r } \boldsymbol{e}_\theta\right).\label{psieq}
\end{equation}
The homogeneous solution of (\ref{psieq}) reads
\begin{equation}
\Psi_h(r,z)=r\int_0^\infty F(k) J_1(kr) e^{-k z} \mathrm{d}k,
\end{equation}
with $F$ to be determined from the boundary conditions. The particular solution can be found from (\ref{omeq}):
\begin{equation}
\boldsymbol{\Omega}=-\nabla^2 \left(\frac{\Psi_p}{r }\boldsymbol{e}_\theta\right)=\frac{\mathrm{Oh}}{\zeta} \nabla^2\boldsymbol{\Omega},
\end{equation}
and hence
\begin{equation}
\Psi_p(r,z)=-r\frac{\mathrm{Oh}}{\zeta}\int_{0}^{\infty}G(k)J_{1}(k r)e^{-z\sqrt{k^{2}+\zeta/\mathrm{Oh}}}\mathrm{d}k.
\end{equation}
Once we know $\Psi=\Psi_h+\Psi_p$, we can find expressions for the velocity field
\begin{eqnarray}
u_{r}(r,z)&=&\int_{0}^{\infty} k F(k)J_{1}(k r)e^{-k z} \mathrm{d}k -\frac{\mathrm{Oh}}{\zeta} \int_{0}^{\infty}\sqrt{k^{2}+\zeta/\mathrm{Oh}} G(k)J_{1}(k r)e^{-z\sqrt{k^{2}+\zeta/\mathrm{Oh}}}\mathrm{d}k, \label{equ_r}\nonumber\\
u_{z}(r,z)&=&\int_{0}^{\infty}k F(k) J_0(k r)e^{-k z} \mathrm{d}k -\frac{\mathrm{Oh}}{\zeta} \int_{0}^{\infty}k G(k)J_{0}(k r)e^{-z\sqrt{k^{2}+\zeta/\mathrm{Oh}}}\mathrm{d}k. \label{equ_z}
\end{eqnarray}
Expressions for $F$ and $G$ can now be obtained from (\ref{bcp2}), (\ref{bcp1}), (\ref{bcp3a}), and (\ref{bc3})
\begin{eqnarray}
& &F(k)-\frac{\mathrm{Oh}}{\zeta}G(k)=\zeta v (k),\nonumber\\
& &k F(k)-\frac{\mathrm{Oh}}{\zeta}G(k)\sqrt{k^2+\zeta/\mathrm{Oh}}=0,\nonumber
\end{eqnarray}
with $v(k)$ given by (\ref{uk}), which give
\begin{eqnarray}
G(k) &=& v (k)\left(k^2+k\sqrt{k^2+\zeta/\mathrm{Oh}}\right),\\
F(k)&=&v(k)\left[\zeta+\frac{\mathrm{Oh}}{\zeta}\left(k^2+k\sqrt{k^2+\zeta/\mathrm{Oh}}\right)\right].
\end{eqnarray}
The pressure in the liquid can now be obtained from (\ref{stokes}). Using (\ref{ustream}) and (\ref{omeq}) one finds that
\begin{equation}
\nabla p=-\zeta \left(\nabla \times \frac{\Psi_h}{r}\boldsymbol{e}_\theta\right),
\end{equation}
 and hence the liquid pressure reads
\begin{equation}
p(r,z)=\zeta\int_{0}^{\infty}F(k)J_{0}(k r)e^{-k z}\mathrm{d}k.
\end{equation}
This results evaluated at $z=0$ is then substituted into (\ref{dyncondstok}) to find (\ref{evpstokes1}) by integration.
Substitution of the Fourier-Bessel series (\ref{fbes}) results in (\ref{evpstokes}).

Due to the complexity of the integral in (\ref{evpstokes}), the system has to be solved iteratively. To this end, we split the integral into parts.  The resulting equation is given by (\ref{eigvstokesht}) with $f$ given by (\ref{fjk}), $g$ by
\begin{eqnarray}
g(j_k,j_n)&=&\int_0^\infty \frac{s^2 J_0^2(s)}{(s^2-j_k^2)(s^2-j_n^2)}\mathrm{d}s \nonumber\\
&=&\left\{ {\begin{array}{ccc}
\frac{4}{\pi}\frac{1}{j_k^2-j_n^2}\left[j_n^2{_2F_3}(1,1;\tfrac{3}{2},\tfrac{3}{2},\tfrac{3}{2};-j_n^2)-j_k^2{_2F_3}(1,1;\tfrac{3}{2},\tfrac{3}{2},\tfrac{3}{2};-j_k^2)\right],&\mathrm{if}& k\neq n,\\
-\frac{4}{\pi}{_2F_3}(1,2;\tfrac{3}{2},\tfrac{3}{2},\tfrac{3}{2};-j_k^2),&\mathrm{if}& k= n.\label{gjk}
\end{array} } \right.,
\end{eqnarray}
and $h$ by
\begin{equation}
h(j_k,j_n, \zeta,\mathrm{Oh})=\int_{0}^{\infty} \frac{ s J_0^2(s ) }{(j_k^2 -s^2) (j_n^2 -s^2)}  \sqrt{s^{2}+\zeta/\mathrm{Oh}} \mathrm{d}s.\label{inth}
\end{equation}

\begin{figure}
	\includegraphics[width=6 in]{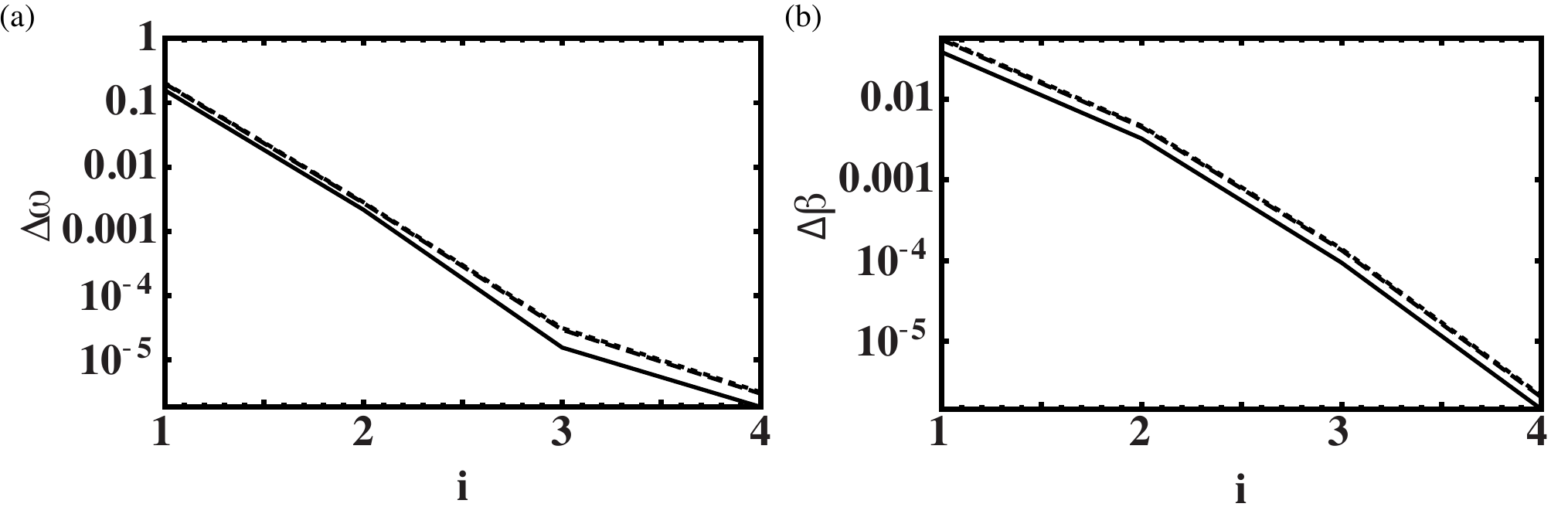}
\caption{(a) The increment in angular frequency $\omega$ of mode 0 versus iteration number $i$ illustrating convergence of the solution process. (b) Increment of damping $\beta$ versus iteration number $i$. The solution is shown for  different matrix sizes: 1x1 (solid) 2x2 (dashed) 3x3 (dotted). }\label{convergence}
\end{figure}

After truncation of the Fourier-Bessel series to $K$ terms, the resulting equation (\ref{eigvstokesht}) in matrix form becomes
\begin{equation}
\left\{\left(\zeta^{(i+1)}\right)^2 \mathsf{A}+2P\mathsf{B}+\frac{1}{2}\mathsf{C}+\zeta^{(i+1)}\mathrm{Oh}\left[\mathsf{G}+\mathsf{H}(\zeta^{(i)},\mathrm{Oh})\right]\right\}\mathbf{c}^{(i+1)}=\mathbf{0},\label{systs}
\end{equation}
with $i$ the iteration number, $\mathsf{A}$ given by (\ref{acoef}), $\mathsf{B}$ given by (\ref{bcoef}), and $\mathsf{C}$ given by (\ref{ccoef});
$\mathsf{G}$ is a $K\times K$-matrix with coefficients 
\begin{equation}
G_{kn}=J_1(j_k)  J_1(j_n)j_kj_n g(j_k,j_n),\label{gcoef}
\end{equation}
and $\mathsf{H}$ is a $K\times K$-matrix with coefficients 
\begin{equation}
H_{kn}=J_1(j_k)  J_1(j_n)j_kj_n h(j_k,j_n,\zeta^{(i)},\mathrm{Oh}).\label{hcoef}
\end{equation}
Figure \ref{convergence} shows that for mode 0, convergence is reached within 4 iterations, irrespective of the matrix size.

\end{document}